\documentclass[a4paper,11pt]{article}
\pdfoutput=1 

\usepackage{jinstpub} 
\usepackage{newtxtext,newtxmath,amsmath}
\usepackage{siunitx}
\title{\boldmath A $ 10^{-3} $ drift velocity monitoring  chamber}
\usepackage{float}

\emailAdd{federica.cuna@le.infn.it, giovanni.tassielli@le.infn.it}

\author[a,b,1]{F. Cuna \note{Corresponding author.}}
\author[d]{, G. Chiarello}
\author[a]{, A. Corvaglia}
\author[e,f]{, N. De Filippis}
\author[a]{, F. Grancagnolo}
\author[b]{, M. Manta}
\author[c,e]{, I. Margjeka}
\author[a]{, A. Miccoli}
\author[a,b]{, M. Panareo}
\author[a,1]{, G. F. Tassielli}
\affiliation[a]{Istituto Nazionale di Fisica Nucleare, Lecce, Italy}
\affiliation[b]{Universit\`{a} del Salento, Italy}
\affiliation[c]{Universit\`{a} degli Studi di Bari, "Aldo Moro", Italy}
\affiliation[d]{Istituto Nazionale di Fisica Nucleare, Roma, Italy}
\affiliation[e]{Istituto Nazionale di Fisica Nucleare, Bari, Italy}
\affiliation[f]{Politecnico di Bari}

\abstract{\noindent The MEG-II experiment searches for the lepton-flavor-violating decay:
	$ \mu\longrightarrow e+\gamma $. The reconstruction of the positron trajectory uses a cylindrical drift chamber operated
	with a mixture of He and $ iC_{4}H_{10} $ gas.
	It is important to provide a stable performance of the detector in terms of its electron transport parameters, avalanche multiplication, composition and purity of the gas mixture. In order to have a continuous monitoring of the quality of gas, we plan to install a small drift chamber, with a simple geometry that allows to measure very precisely the electron drift velocity in a prompt way. This monitoring chamber will be supplied with gas coming from the inlet and the outlet of the detector to determine if gas contaminations originate inside the main chamber or in the gas supply system. The chamber is a small box with cathode walls, that determine a highly uniform electric field inside two adjacent drift cells. Along the axis separating the two drift cells, four staggered sense wires alternated with five guard wires collect the drifting electrons. The trigger is provided by two $ ^{90}Sr $ weak calibration radioactive sources placed on top of a two thin scintillator tiles telescope. The whole system is designed to give a prompt response (within a minute) about drift velocity variations at the $ 10^{-3} $ level.
	}

\keywords{Drift chambers, Particle tracking detectors, Gaseous detectors, Models and simulations, Electric fields, Charge transport and multiplication in gas}

\proceeding{International Conference on Instrumentation for Colliding Beam Physics\\
 24 - 28 February, 2020\\
 Budker Institute of Nuclear Physics, Novosibirsk, Russia }

\begin{document}
\maketitle
\flushbottom

\section{Introduction: motivation for a drift velocity monitoring chamber}
\label{sec:intro}

The choice of a gas mixture in a drift chamber is of utmost importance, in particular for the experiments, like MEG-II, in which the trajectories of low momentum particles need to be reconstructed with high accuracy. Moreover, it is crucial to control the purity of gas injected in the drift chamber because uncontrolled fluctuations of the gas composition and
contaminations by impurities would make the drift velocity unstable and could deteriorate spatial and momentum resolution of candidate signal tracks.\\
Several studies about the behaviour of drift velocity as a function of the reduced electric field in a mixture of $He/ iC_{4}H_{10} $ have been published.	
\begin{figure}[H]
	\centering 
	\includegraphics[width=6cm,height=6cm]{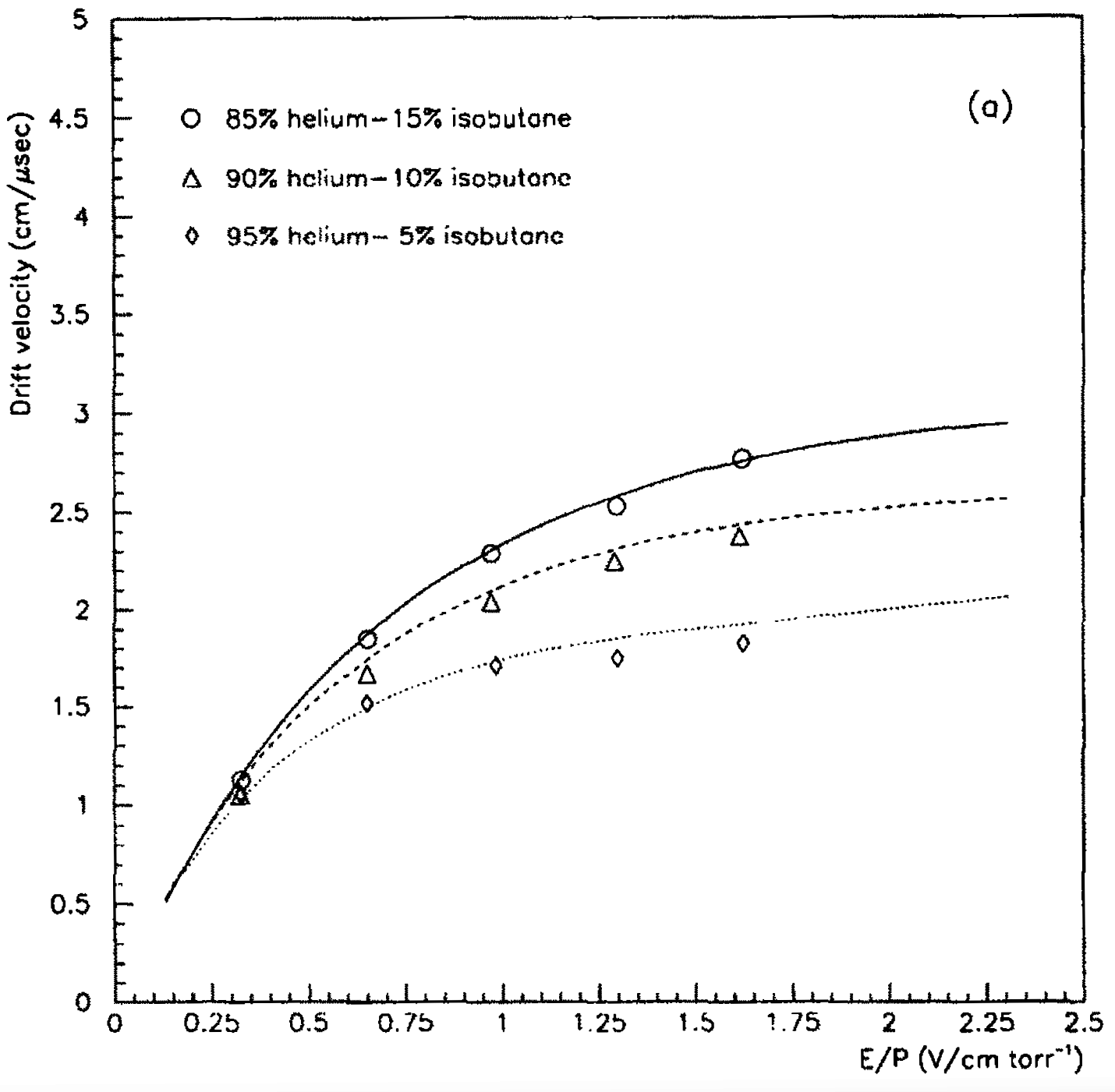}\quad
	\includegraphics[width=6.3cm,height=6cm]{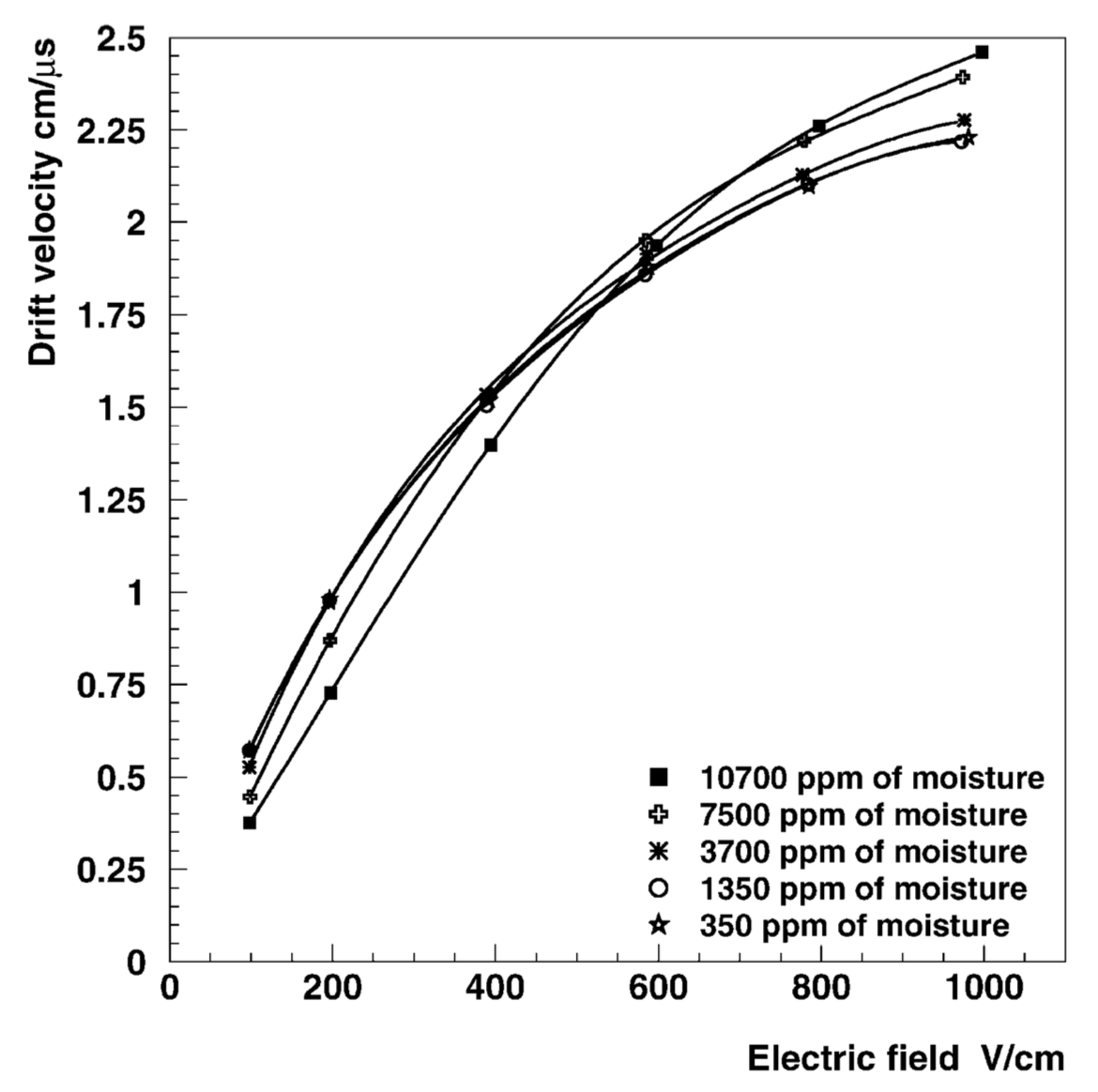}
	\caption{Left, drift velocity as a function of the reduced electric field for different percentages of helium-isobutane mixtures \cite{1} and right, drift velocity as a function of the applied electric field for different concentration of water vapours \cite{2}.}
	\label{drift_trend}
\end{figure}
\noindent 
Drift velocity is the most sensitive parameter for the operation of a drift chamber with respect to tiny variations of the gas mixture and so it is the "target" that we want to use in order to control the gas purity. As an example, the left graph in Figure \ref{drift_trend} shows that for a mixture of $He/ iC_{4}H_{10} $ at normal pressure, variations of the electric field, around the operating value of 1 V/cm $torr^{-1}$, of about 2 $ V/cm$ induce drift velocity variations of about $ 1\times10^{-3} $.\\
Moreover, to mitigate the ageing effect, it is useful to introduce small quantities of water vapors in the gas mixtures, but it is mandatory to control the consequent variations of drift velocity. As an example, the right graph of Figure \ref{drift_trend} shows that at the operating value of the electric field of about 1 V/cm $torr^{-1}$, variations of $ \approx $ 150 ppm lead to an increase up to $1 \times 10^{-3} $ in drift velocity.
\section{ Monitoring drift chamber }
\begin{figure}[H]
	\centering
	\includegraphics[width=0.7\textwidth]{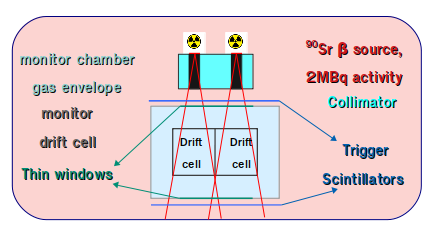}
	\caption{Experimental set up of the monitoring drift chamber.}
	\label{setup}
\end{figure}

\begin{figure}[H]
	\centering
	\includegraphics[width=0.6\textwidth]{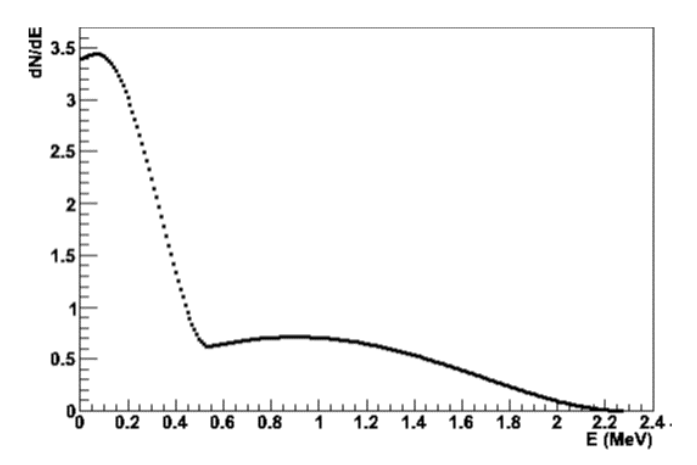}
	\caption{Energy spectrum of electrons emitted by the $ ^{90}Sr $ radioactive source.}
	\label{Spectrum}
\end{figure}

The main goal of the monitoring chamber is to provide a prompt response about drift velocity variations at $ 10^{-3} $ level.\\
This purpose can be obtained with a conceptually very simple structure, illustrated in Figure \ref{setup}.
\noindent
We will use two $^{90}Sr $ low-activity calibration radioactive sources placed on top of two thin scintillator tiles telescope. The sources will be collimated to select the tracks crossing the drift cells.\\
The  $ ^{90}Sr $ sources produce $ 2\times10^{6}$ electrons per second, with the energy distribution shown in Figure \ref{Spectrum}. Only the electrons with an energy larger than approximately 0.8 MeV, amounting to about $20\%$ of the total \cite{3}, will be able to cross the chamber and trigger the scintillator telescope. 
\noindent
 Moreover, considering the solid angle acceptance, purposedly defined by the source collimators, the total number of triggering decay electrons will be around  $ 4\times10^{3} $ . 
\subsection{Mechanical details of the monitoring drift chamber}
\begin{figure}[H]
	\centering
	\includegraphics[width=7.1cm]{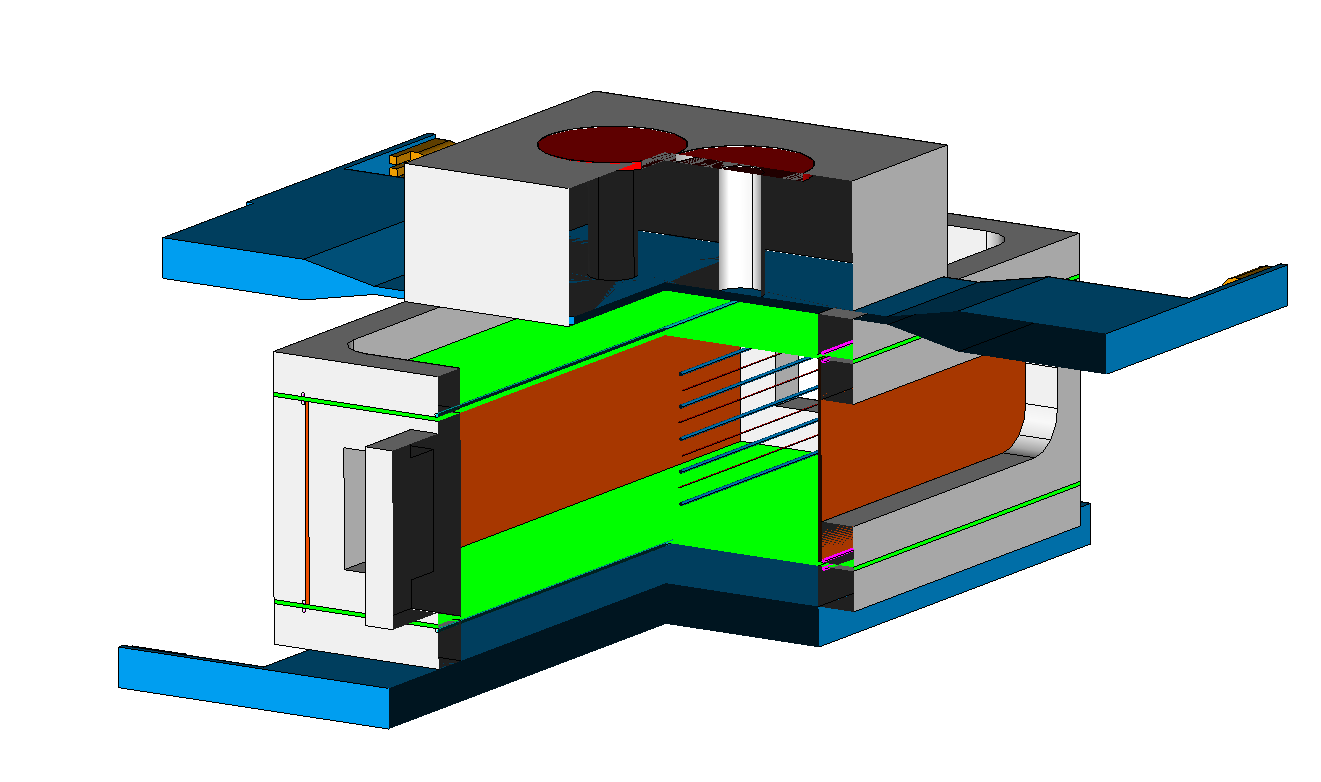}\quad \includegraphics[width=7.6cm]{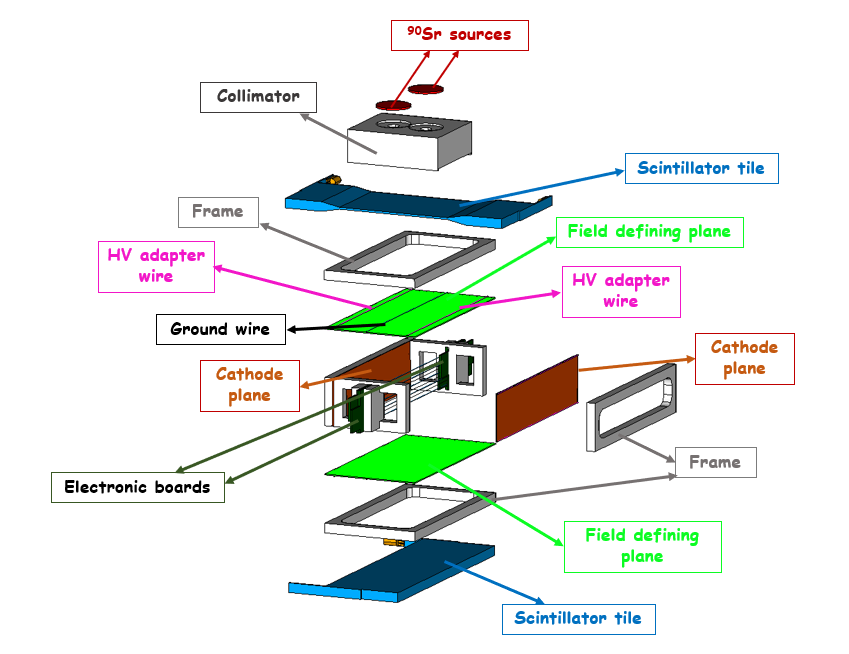}
	\caption{Cross-cut section (left) and exploded view (right) of the mechanical setup.}
	\label{mec2}
\end{figure}

\begin{figure}[H]
	\centering
	\includegraphics[width=0.8\textwidth]{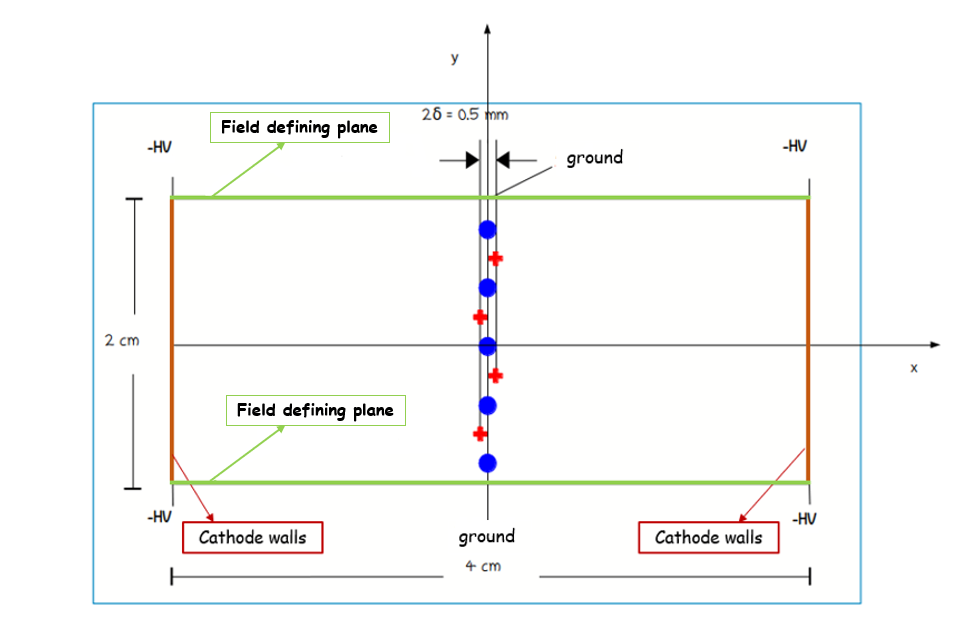}
	\caption{The figure shows sense wires staggering used to measure drift velocity .}
	\label{StagWires}
\end{figure}

The mechanical structure of the monitoring drift chamber is presented in Figure \ref{mec2}.\\
\noindent
The lateral cathode walls are made of a thin (190 $\mu m$) Cu coated PET foil, glued to a rigid frame to preserve planarity. The uniform electric field across the drift cells is obtained with a resistive \\25 $\mu m$ DLC foil, with an electrical resistivity around $100$ $M\Omega$ $m$, connected at the edges to the high voltage lateral cathode walls and with a longitudinal wire in the middle, connected to ground. Along the plane separating the two drift cells, four sense wires (20 $\mu m$ diameter gold plated tungsten) alternated to five guard wires (80 $\mu m$ diameter silver plated aluminum) collect the drift electrons, as Figure \ref{StagWires} shows.
\noindent
Sense wires are staggered in the plane $x=0$ by $\pm\delta=\pm 0.25$ $\mu m$ to allow drift velocity measurements.
\section{Simulation of electric field configuration}
\begin{figure}[h]
	\centering
	\includegraphics[width=0.6\textwidth]{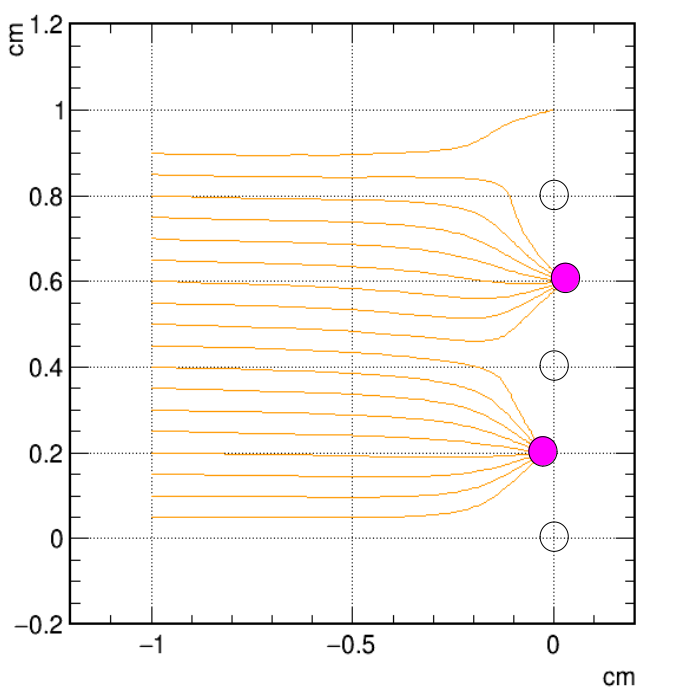}
	\caption{The figure shows the drift lines converging on the sense wires, marked with the pink color.}
	\label{drift_line}
\end{figure}

The drifting electric field and the amplification field around the sense wires have been calculated with the Garfield++ program.
For a drifting field of 1 $kV/cm$ (corresponding to $-2000$ $V$ on the cathode planes) and a gas amplification gain of about $ 5\times10^{5}$ on the sense wires, given the described wire diameters, the voltage on the sense wires must be set at $+1000$ $V$, whereas, on the guard wires it needs to be $ >= -350$ $V$ in order to keep the value of the electric field on the guard wire surface above $-20$ $kV/cm$, thus avoiding amplification of positive ions.\\
Figure \ref{drift_line} shows the electron drifting lines to the sense wires in the described electrostatic configuration. The asymmetry introduced by the sense wire staggering is confined to a very limited region around the wires (i.e. for very short drift times) and, as it will become clear in the next paragraph, its effects will be systematically subtracted in the calculation of the drift velocity.

\section{Sensitivity of drift velocity measurement}
\begin{figure}[H]
	\centering
	\includegraphics[width=0.8\textwidth]{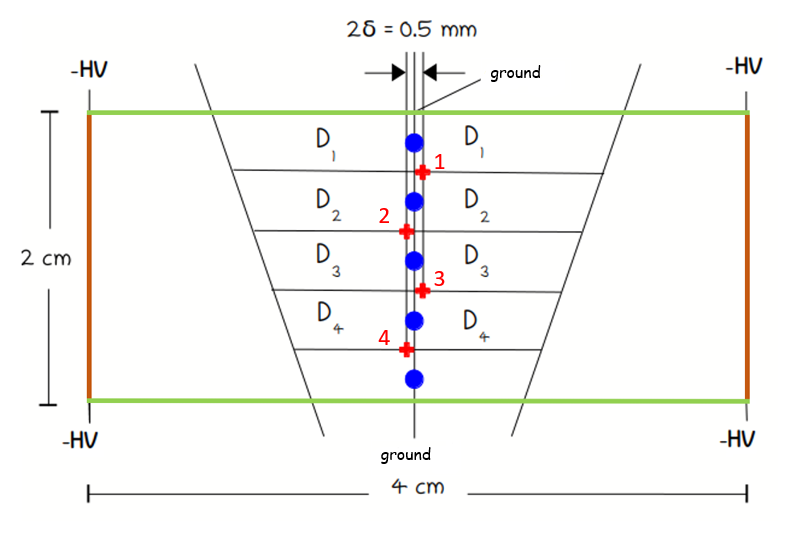}
	\caption{The figure shows drift cells structure and two tracks passing inside them. $D_{i}$ indicate the drift distance from the crossing track to the sense wire i.}
	\label{trackC}
\end{figure} 
Considering the two triplets of the wires (\textit{123}) and (\textit{234}), as shown in Figure \ref{trackC}, indicating with $v_{d}$ the constant drift velocity in the uniform electric field region, simple geometrical considerations lead to the following relations
\begin{align}
t_{2}=\frac{t_{1}+t_{3}}{2}\mp\frac{2\delta}{v_{d}}\\
t_{3}=\frac{t_{2}+t_{4}}{2}\pm\frac{2\delta}{v_{d}}\text{,}
\end{align}
where the first (second) choice of the signs refers to a track crossing the right (left) side drift cell.
Defining the variable:
\begin{equation}
\Theta=(t_{1}+t_{3}-2t_{2})-(t_{2}+t_{4}-2t_{3}) \text{,}
\end{equation}
which, according to the track crossing side, assumes one of the two values:
\begin{equation}
\begin{cases}
\Theta_{+}=+\frac{8\delta}{v_{d}} \quad\quad left\\
\Theta_{-}=-\frac{8\delta}{v_{d}} \quad\quad right 
\end{cases}
\end{equation}
one obtains the estimate of $v_{d}$ and of its variance as a function of $\Delta \Theta = |\Theta_{+} - \Theta_{-}|$ :
\begin{equation}
\begin{aligned}
v_{d}&=\frac{16\delta}{\Delta\varTheta} \\
\sigma_{v_{d}}&=\sqrt{\left(\frac{16}{\Delta\varTheta}\right)^{2}\sigma_{\delta}^{2}+\left(-\frac{16\delta}{\Delta\varTheta^{2}}\right)^{2}\sigma_{\Delta\varTheta}^{2}} \text{,}
\label{v_drelation}
\end{aligned}
\end{equation}
where $ \sigma_{\delta} $ represents the error on the wire positioning and $\sigma_{\Delta\varTheta}$ depends statistically on the number of events collected. Since one is interested only in the variations of the drift velocity, the first contribution cancels out and the precision will scale with the collected statistics.\\

\section{Simulation of the procedure} 
\begin{figure}[H]
	\centering
	\includegraphics[width=1\textwidth]{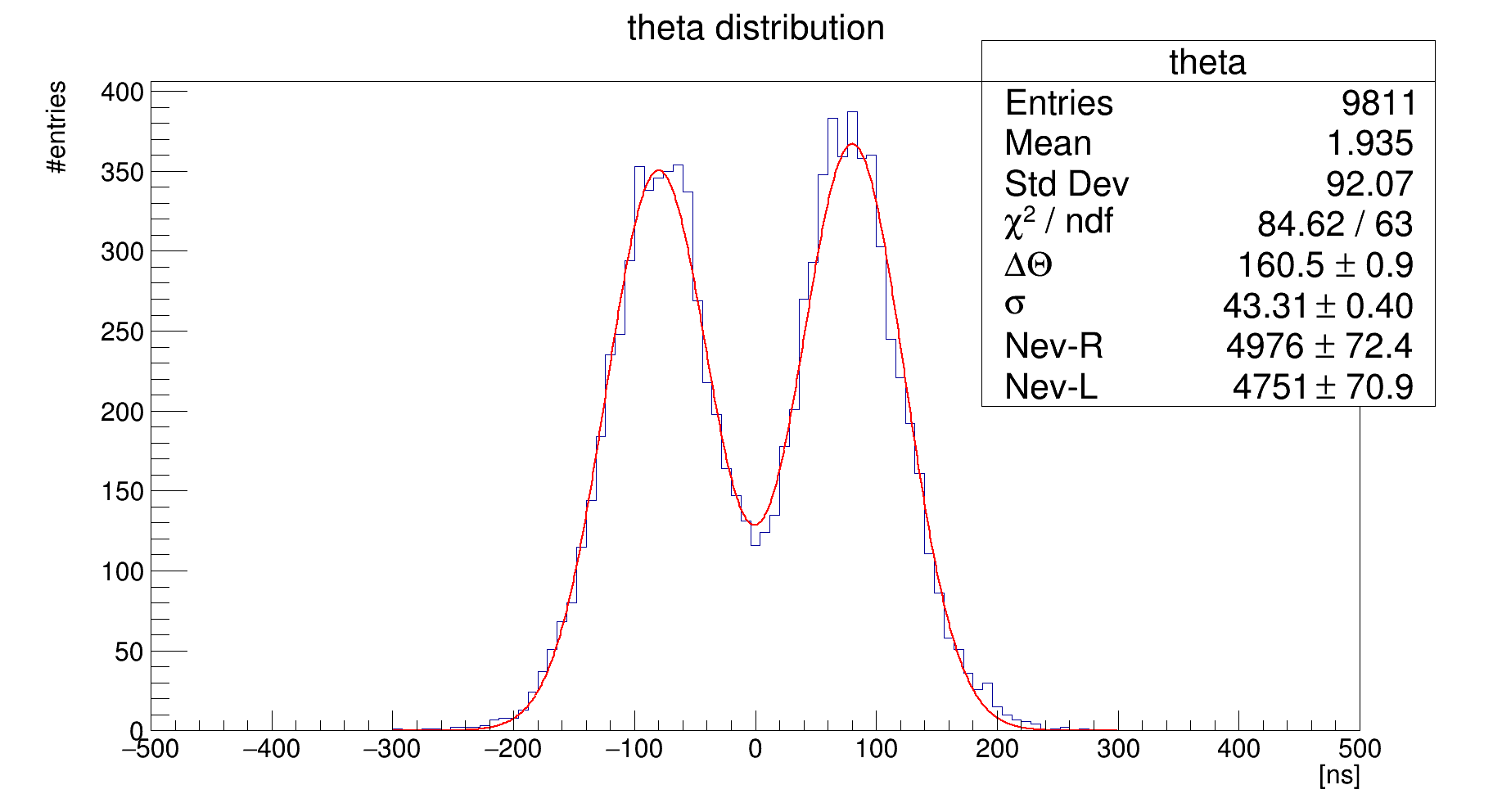}
	\caption{The double peak distribution for a mixture of  $ 90\% He-10\%iC_{4}H_{10} $ at pressure of 760 torr and temperature of 300 K. }
	\label{ex_dpeak}
\end{figure}

For the measurement of the drift velocity, we simulated 9000 tracks, 4500 passing through the left side and 4500 passing through the  right side of the chamber. The gas mixture chosen for the simulation is of $ 90\% He-10\%iC_{4}H_{10} $ at pressure of 760 torr and temperature of 300 K .\\
The tracks are generated with a uniform angular distribution within  $ \pm12^\circ $.\\
 For every track, four drift times are collected and the value of the variable $\Theta$ is plotted in the histogram of Figure \ref{ex_dpeak}. The two peaks, corresponding to $\Theta_{+}$ and $\Theta_{-}$, are highlighted by a fit to the distribution.\\

\noindent
The drift velocity calculated from this distribution is $2.48 \pm 0.02$ $cm/\mu s$, according to the relation \eqref{v_drelation}. The number of the events simulated allows for a sensitivity of $1 \times 10^{-2}$, but an increment of a factor 100 in the number of tracks will give the expected sensitivity of $1 \times 10^{-3}$.
\section{Conclusion}
The monitoring drift chamber, equipped with low-activity radioactive sources and triggered by a telescope of thin scintillator tiles, allows one to monitor the drift velocity of the MEG-II central tracker, in short time and with a high precision that allows one to evaluate variations of the operating conditions, which would affect spatial resolution.\\
The continuous monitoring of drift velocity variations at 1 \textperthousand\ level is sensitive to variations of:
\begin{itemize}
	\item $+0.4\%$ in $iC_4H_{10}$ content (from $10.0\%$ to $10.4\%$)
	\item $-0.2\%$ in in $iC_4H_{10}$ content (from $10.0\%$ to $9.8\%$)
	\item $±0.4\%$ in E/p ($\approx6\%$ in gas gain) at gain $\approx 5\times10^5$
	\item $\mp 4$ $V$ at $p \approx 1$ $bar$, $T\approx$\SI{25}{\celsius}
	\item $\mp 4$ $mbar$ at $V\approx1500$ $V$, $T\approx$\SI{25}{\celsius} 
	\item \SI{-0.3}{\celsius} at $p\approx 1$ $bar$, $V\approx1500$ $V$
	\item $\leq100$ ppm variations in water vapor content around $3500$ ppm
\end{itemize}



%
%


%
%
%
%
%
%
%
%

\begin{thebibliography}{99} 
	
	\bibitem{1}P. Bernardini, G. Fiore, R. Gerardi, F. Grancagnolo, U. von Hagel , F. Monittola, V. Nassisi, C. Pinto, L. Pastore, M. Primavera, \emph{Precise measurements of drift velocities in helium gas mixtures}, 
 	\href{https://doi.org/10.1016/0168-9002(94)01144-3}	{\emph{Nuclear Instruments and Methods in Physics Research} {\bf A} 355 (1995) 428-433}
	
	\bibitem{2} V. Golovatyuk, F. Grancagnolo, R. Perrino, \emph{Influence of oxygen and moisture content on electron life time in helium-isobutane gas mixtures},
	\href{https://doi.org/10.1016/S0168-9002(00)01172-4}{\emph{Nuclear Instruments and Methods in Physics Research} {\bf A} 461 (2001) 77-79.}

	\bibitem{3} J.J. Devaney, \emph{Beta spectra of $^{90}Sr$ and $^{90}Y $},
	\href{https://www.osti.gov/servlets/purl/5143758-BtuIGS/}{\emph{United States: N. p.} (1985).}
	

	
	
\end{thebibliography}
\end{document}